\documentstyle[aps,prl,multicol,epsf]{revtex}
\renewcommand{\narrowtext}{\begin{multicols}{2} \global\columnwidth20.5pc}
\renewcommand{\widetext}{\end{multicols} \global\columnwidth42.5pc}
\renewcommand{\v}[1]{{\bf #1}}

\newcommand{\rhob}{{\bar{\rho}}}

\newcommand{\phib}{{\bar{\phi}}}

\newcommand{\w}{{\omega}}

\newcommand{\ba}{\begin{eqnarray}}
\newcommand{\ea}{\end{eqnarray}}
\newcommand{\be}{\begin{equation}}
\newcommand{\ee}{\end{equation}}
\newcommand{\nn}{\nonumber\\}
\newcommand{\Eq}[1]{Eq.~(\ref{#1})}
\newcommand{\p}{\partial}

\newcommand{\ra}{\rightarrow}

\newcommand{\cL}{ {\cal L} }

\begin{document}
\draft
\title{A Duality Between Unidirectional Charge Density Wave Order and Superconductivity}
\author{Dung-Hai Lee}
\address{Department of Physics,University of California
at Berkeley, Berkeley, CA 94720, USA \\} \maketitle \draft
\begin{abstract}
This paper shows the existence of a duality between an  unidirectional
charge density wave order and a superconducting order.
This duality predicts the existence of charge density wave near a superconducting vortex, and the
existence of superconductivity near a
charge density wave dislocation.
\end{abstract}
\narrowtext

It is well known that depending on the doping level the high $T_c$ superconductors exhibit antiferromagnetism and
d-wave superconductivity. Tranquada {\it et al} showed that for the high $T_c$ compound $La_{2-x}Sr_xCuO_4$, four
incommensurate neutron peaks appear around $(\pi,\pi)$ when appropriate amount of $Sr$ are replaced by
$Nd$\cite{nd}. The positions of these peaks are consistent with the earlier theoretical proposal of an
unidirectional spin density wave, or stripe, order\cite{stripe}.

Recently Lake {\it et al} showed that in the highest $T_c$ LSCO a moderate magnetic field can considerably lower the
energy of the incommensurate peaks\cite{lake}. One interpretation
of this result is that the stripe order closely competes with superconductivity
so that a small suppression of the latter by magnetic field allows
the former to appear\cite{so5,app}. Based on this interpretation a number of interesting theoretical works
appear recently. In these works the LSCO system is viewed as having two competing order parameters - the
stripe order and superconductivity\cite{app}.

The purpose of the present work is to extend the Landau-like theories in Ref.\cite{app}to
include quantum fluctuations of the stripe and superconducting order
parameters\cite{ed}. This theory is relevant when the Landau theory predicts the coexistence of both orders.
A caveat of the present theory is that it assumes an unidirectional charge density
wave instead of spin density wave order. For this reason this
theory is not obviously applicable to the experiments reported in
Ref.\cite{lake}.

Although this theory is motivated by the ``stripe phenomena'' in high $T_c$, it is should be perceived more
generally as a quantum theory for competing superconducting and unidirectional charge density wave orders. Since
this theory is phenomenological in nature it sheds no light on the superconducting pairing mechanism and the
origin of stripe order. Our main results are summarized in the last paragraph of this paper.

The basic assumption we make in this work is the coexistence of the stripe and superconducting orders in the
mean-field theory. Our goal is to go beyond mean-field and describe the soft collective fluctuations. Unless
stated otherwise the entire discussion assumes zero temperature. The effective actions presented in this work are
the low-energy and long wavelength quantum Boltzmann weight in 2 space and 1 (imaginary) time dimensions. The
space cutoff of these actions is the maximum of the superconducting paring length and the distance between
neighboring stripes in the mean-field theory.

We begin with the theory for superconducting phase fluctuations\cite{lf}. When the modulus of the superconducting
order parameter $\psi_{sc}$ is well-developed, the low-energy and long-wavelength excitations are associated with
the fluctuations of the superconducting phase $\theta_s$ ($\psi_{sc}=|\psi_{sc}|e^{i\theta_s}$). Under that
condition the following U(1) sigma-model describes the physics at long distance and low energy \ba
\cL_{\phi}&&=\sum_{\mu=x,y,t}\frac{K_{\phi \mu}}{2}|\p_{\mu}\phi|^2+ R_{\phi}\phib\p_t\phi.\label{phi}\ea In the
above $\phi=e^{i\theta_s}$ is a U(1) phase factor. The periodicity of $\phi$ under $\theta_s\ra\theta_s+2 n\pi$ is
crucial for the existence of vortices. The last term of the above equation is commonly referred to as the Berry's
phase\cite{lf}.  $R_{\phi}$ acts as a chemical potential that drives the Cooper pair density away from
integral number per unit cell. 
By introducing auxiliary fields $J_x,J_y,J_t$ we can Hubbard-Stratonavich transform \Eq{phi} and rewrite it as
 \ba \cL_{\phi}&&=\sum_{\mu}[\frac{1}{2K_{\phi \mu}}J_{\mu}^2+ J_{\mu}\phib\p_{\mu}\phi
]+R_{\phi}\phib\p_t\phi.\label{d1}\ea We note that after integrating out $J_{\mu}$ and using the identity
$(\phib\p_{\mu}\phi)^2=-|\p_{\mu}\phi|^2$ \Eq{phi} is recovered.

If we separate $\theta_s$ into a part with winding (i.e vorticity) and a part without, and integrate out the
latter we obtain the continuity equation\cite{lf} $\sum_{\mu} \p_{\mu}J_{\mu}=0$. Physically $J_{\mu}$ is the
3-current of the Cooper pairs. Due to the continuity equation we can find a ``gauge field'' $a_{\mu}$ so that $
J_{\mu}=\sum_{\nu\lambda}\epsilon_{\mu\nu\lambda}\p_{\nu}a_{\lambda}.$ Substitute the above equation into \Eq{d1}
we obtain \be  \cL_{\phi}=\sum_{\mu}[\frac{1}{2K_{\phi \mu}}f_{\mu}^2+ i ( a_{ex,\mu}+a_{\mu})V_{\mu}].\label{dd1}
\ee In the above $f_{\mu}=\sum_{\nu\lambda}\epsilon_{\mu\nu\lambda}\p_{\nu}a_{\lambda}$ is the dual field strength
of $a_{\lambda}$, $a_{ex,\mu}=(0, a_{ex,1},a_{ex,2})$ with $\p_1a_{ex,2}-\p_2 a_{ex,1}=R_{\phi}$, and
$V_{\mu}=-i\sum_{\nu\lambda}\epsilon_{\mu\nu\lambda}\p_{\nu}\phib\p_{\lambda}\phi$ is the vortex
3-current\cite{lf}. \Eq{dd1} describes the physics of \Eq{phi} from the point of view of vortices. The Cooper-pair
current $J_{\mu}$ becomes the electric and magnetic fields that couple to the (bosonic) vortex current $V_{\mu}$
minimally,  $R_{\phi}$ becomes a background magnetic field.

\Eq{phi} involves four parameters $K_{\phi 0,1,2}$ and $R_{\phi}$. When $R_{\phi}=0$ \Eq{phi} is the anisotropic 3D xy model. Depending on the strength of $K_{\phi\mu}$ the system
can be superconducting or insulating. In the superconducting phase the vortex current $V_{\mu}$ are absent at low
energy/long wavelength. Thus one can view the superconductor as a vortex insulator. In the insulating phase the
vortex current fluctuates severely (i.e. vortices become superconducting) which quenches the Cooper current
fluctuation. Thus the Cooper pair insulator is a vortex superconductor. When $R_{\phi}\ne 0$ the vortex
condensation is frustrated by a background magnetic field. In that case \Eq{phi} describes a superconductor.

Next we turn to the theory for fluctuating unidirectional charge density wave. A review of the classical
statistical mechanics of fluctuating density wave order parameters can be found in Ref.\cite{nel}. In the
following we generalize the treatment of Ref.\cite{nel} to include the imaginary time dimension.

For concreteness let us assume the stripe order causes the charge density to modulate at an incommensurate
wavevector $\v G=G \hat{x}$.
The Fourier component of the charge density at this wavevector serves as the order parameter. Since we assume the
breaking of rotational symmetry the ``star'' of the ordering wavevectors consists of a single pair $\v G$ and $-\v
G$. This leads to a single complex number $\psi_{cdw}$ (and its complex conjugate $\psi_{cdw}^*$) as the order
parameter. We further assume the modulus of $\psi_{cdw}$ (which is a complex scalar\cite{nel}) is well developed.
The low energy excitations are dominated by the fluctuations in the phase $\theta_{\rho}$ of $\psi_{cdw}$ (i.e.
$\psi_{cdw}=|\psi_{cdw}| e^{i\theta_{\rho}}$). Physically the $\theta_{\rho}$ fluctuation is due to the
displacement of stripes from their ideal position. If $\v u (\v r,t)$ is the displacement field, the relation
between $\theta_{\rho}$ and $\v u$ is $\theta_{\rho}=\v G\cdot\v u$.

Let us define $\rho\equiv e^{i\theta_{\rho}}$,  the low energy and long wavelength stripe dynamics is described by
the following effective Lagrangian \ba \cL_{\rho}=\sum_{\mu=x,y,t}\frac{K_{\rho
\mu}}{2}|\p_{\mu}\rho|^2.\label{rho}\ea The $\mu=x,y$ terms of the above equation are the elastic energy due to
the stripe displacement, and the $\mu=t$ term is the associated kinetic energy.

Similar to the passage from \Eq{phi} to \Eq{d1} we Hubbard-Stratonavich transform \Eq{rho} by introducing
auxiliary fields $q_{\mu}$ \ba \cL_{\rho}=\sum_{\mu}[\frac{1}{2K_{\rho \mu}}q_{\mu}^2+
q_{\mu}\rhob\p_{\mu}\rho].\label{d2}\ea  By separating $\theta_{\rho}$ into a part involves dislocations and a
part does not and integrating out the latter we obtain $\sum_{\mu}\p_{\mu}q_{\mu}=0.$ Physically $q_{0}$ is the
momentum density and $q_{1,2}$ are the components of the momentum current associated with the stripe displacement.
Due to the conservation of $q_{\mu}$ we can find a ``gauge field'' $b_{\mu}$ so that $
q_{\mu}=\epsilon_{\mu\nu\lambda}\p_{\nu}b_{\lambda}.$ Substituting the above into \Eq{rho} we obtain
\be
\cL_{\rho}=\sum_{\mu}[\frac{1}{2K_{\rho \mu}}\tilde{f}_{\mu}^2+ i b_{\mu}D_{\mu}].\label{ddd1} \ee In the above
$\tilde{f}_{\mu}$ is the dual field strength of $b_{\mu}$ and $
D_{\mu}=-i\epsilon_{\mu\nu\lambda}\p_{\nu}\rhob\p_{\lambda}\rho$ is the 3-current of the bosonic dislocations.
Here a few words should be said about $D_{\mu}$. In general $D_{0}$ for a {\it two-dimensional} charge density
wave should be a vector. The fact that $D_0$ in the present theory is a scalar is due to the one-dimensional
nature of the charge density wave.

Depending on the strength of $K_{\rho\mu}$ \Eq{rho} can describe the stripe-ordered or the stripe-disordered
phases. In the stripe-ordered phase the fluctuations of $D_{\mu}$ are absent at low energy and long wavelength,
while in the stripe disordered phase $q_{\mu}$ fluctuation is absent. Thus the stripe ordered phase is an
insulator of the dislocations, while the quantum disordered state of stripe can be viewed as a Bose condensate of
dislocations.

In the above we assumed that the charge density wave ordering wavevector $\v G$ is incommensurate with the
reciprocal lattice vector of the underlying lattice. When $\v G$ is commensurate with the reciprocal lattice
vector pinning becomes an important issue. The lowest order effect of pinning is to exert the following potential on the
displacement field $\v u(\v r)$ (recall $\theta_{\rho}(\v r)=\v G\cdot\v u(\v r)$) \be
L_{pin}=-\Gamma\sum_i\cos{(G x_i+\theta_{\rho}(\v r_i))}.\label{pin} \ee Here $i$ label the lattice site and $\v
r_i$ is the corresponding position vector. When the stripe period is $p/q$ times the lattice constant $Gx_i=(2\pi
q/p) m_i$ where $m_i$ is an integer between $0$ and $p-1$. Unless $p=q=1$ this site-dependent phase leads to cancellation
when we
perform the lattice sum. The lowest order pinning term that survives the lattice
summation is \be L_{pin}'=-\Gamma'\sum_i \cos{(p \theta_{\rho}(\v r_i))}.\ee Strictly speaking such term
always pin the stripe at zero temperature (in two space dimensions)\cite{monopole}. However it is important to
bear in mind that for large $p$ the pinning strength $\Gamma'\sim \Gamma^P$ can be very small, and hence  the
pinning is only important at extremely low energies. In that case for practical purpose we can ignore
the pinning.

Having understood the effective theories for the superconducting and the stripe orders separately, we are ready to
treat them simultaneously. The most important question is how to coupled theses two order parameters together. To
answer this question we take hints from experiments. Empirically it is known that changing the doping density
affects the stripe period. In particular for sufficiently low doping the stripe period is inversely proportional
to the doping density\cite{yamada}. This suggests that a non-zero $J_t$ tends to shift the phase of $\rho$, i.e.
$\rho\ra e^{ig_1 J_t x}\rho,$ which in turn means a change of the stripe ordering wavevector from $G\hat{x}$ to
$(G+g_1 J_t)\hat{x}$. Similarly a non-zero current $J_x$ tends to boost the stripes and shift the phase of $\rho$
according to $\rho\ra e^{ig_2 J_x t}\rho.$ This shift changes the stripe displacement in a time-dependent fashion:
$\v u(\v r,t)\ra \v u+g_2 J_x t \hat{x}$. These considerations suggest that \Eq{d1} and \Eq{rho} be coupled
together as follow \ba \cL&&=\frac{K_{\rho x}}{2}|(\p_x-ig_1J_t)\rho|^2+ \frac{K_{\rho
t}}{2}|(\p_t-ig_2J_x)\rho|^2\nn&&+\frac{K_{\rho y}}{2}|\p_y\rho|^2+\sum_{\mu}[\frac{1}{2K_{\phi \mu}}J_{\mu}^2 +
J_{\mu}\phib\p_{\mu}\phi ] +R_{\phi}\phib\p_t\phi.\label{tot}\ea We note that in \Eq{tot} $J_t$ and $ J_x$ act
like the x and t components of a ``gauge field'' coupling minimally to the gradient of the stripe order parameter,
suggesting that stripe order tends to suppress the fluctuation in $J_{x,t}$.

Analogously to the passage from \Eq{rho} to \Eq{d2} we introduce auxiliary fields $q_{\mu}$ to
Hubbard-Stratonavich transform the first three terms of \Eq{tot}. The result is \ba
\cL&&=\sum_{\mu}[\frac{1}{2K_{\rho \mu}}q_{\mu}^2+\frac{1}{2K_{\phi \mu}}J_{\mu}^2 + J_{\mu}\phib\p_{\mu}\phi
]+R_{\phi}\phib\p_t\phi\nn&&+q_x(\rhob\p_x\rho-ig_1J_t)+ q_y\rhob\p_y\rho +
q_t(\rhob\p_t\rho-ig_2J_x).\label{totd1}\ea Integrate out $J_{\mu}$ in the above equation we obtain \ba
\cL&&=\frac{K_{\phi x}}{2}|(\p_x-ig_2q_t)\phi|^2+ \frac{K_{\phi t}}{2}|(\p_t-ig_1q_x)\phi|^2\nn&&+\frac{K_{\phi
y}}{2}|\p_y\phi|^2+\sum_{\mu}[\frac{1}{2K_{\rho \mu}}q_{\mu}^2 + q_{\mu}\rhob\p_{\mu}\rho ] +
R_{\phi}\phib\p_t\phi.\label{totd2}\ea Comparing \Eq{totd2} with \Eq{tot} we see that aside from the last term
these two equations can be made identical after the following replacements \ba \rho&&\leftrightarrow\phi, ~~~
K_{\phi\mu}\leftrightarrow K_{\rho\mu}\nn q_{\mu}&&\leftrightarrow J_{\mu}, ~~~ g_1\leftrightarrow
g_2.\label{rep}\ea \Eq{rep} is the main result of this paper, it implies that when $R_{\phi}=0$ the stripe order
and the superconducting order are dual to each other.

\Eq{tot} generically supports four different phases : a) $<\phi>\ne 0, <\rho>\ne 0$ ; b) $<\phi>\ne 0, <\rho>=0$;
 c) $<\phi>=0, <\rho>\ne 0$ ; and d) $<\phi>=0, <\rho>=0$.
Since both disordered phase is not terribly interesting, in the rest of the paper we shall concentrate on a)-c).

Phase a) 
has both $\phi$ and $\rho$ ordered. (It is important to note that by $<\phi>\ne 0, <\rho>\ne 0$ we refer to the
coexistence of the superconducting and stripe order above the cutoff length, i.e., the stripe period. It does not
imply the homogeneous coexistence of these two types of order at microscopic scale.) In this phase we can ignore
the vortices in $\phi$ and the dislocation in $\rho$ and arrive at a quadratic theory with two linearly dispersive
gapless normal modes. It is straightforward to compute the density-density and current-current correlation
functions using this quadratic
theory. 
The results are summarized as follows. \ba &&\kappa_{\w}=\frac{K_{\phi x}K_{\phi y}}{K_{\phi
x}\cos^2{\theta}+K_{\phi y}\sin^2{\theta}}\nn&& \kappa_{q}=\frac{K_{\phi x}K_{\phi y}}{K_{\phi
x}\cos^2{\theta}+(K_{\phi y}+g_1^2K_{\phi x}K_{\phi y}K_{\rho t}) \sin^2{\theta}}\nn&&\pi_q=K_{\phi
t}\frac{K_{\rho x}\cos^2{\theta}+K_{\rho y}\sin^2{\theta}}{K_{\rho x}\cos^2{\theta}+(K_{\rho y}+g_2^2K_{\rho
x}K_{\rho y}K_{\phi t})\sin^2{\theta}}.\ea  Here $\kappa_{w}$ is the transverse current-current correlation
function in the limit $\v q\ra 0$ and $\w\ra 0$ while $|\v q|/\w\ra 0$. Physically it gives the superfluid density
measured in optical conductivity. $\kappa_q$ is the current-current correlation function in the limit $\v q\ra 0$
and $\w\ra 0$ while $\w/|\v q|\ra 0$. Physically it gives the the superfluid density in the DC magnetization
measurement. Finally $\pi_{q}$ is the DC density-density correlation function.

Due to the anisotropy the response function depends on the angle $\theta$ between the vanishing wavevector and
$\hat{x}$ (the stripe ordering direction).
We note that in the presence of stripes $\kappa_{\w}\ne \kappa_q$. In principle this inequality can be seen by
comparing the optical conductivity result with the finite wavelength DC magnetization measurement. Despite this
slight complication the above results imply that phase a) is an anisotropic superconductor.

Next we focus on phase b) where there is superconductivity but no stripe order. In this case this theory has an
interesting implication on the $\rho$ order in the neighborhood of a superconducting vortex. Around a
superconducting vortex the phase of $\phi$ winds by $2\pi$. In order to minimize the gradient energy in the first
term of \Eq{totd2} a non-zero $q_t$ is induced: \be <q_t(\v x,t)>\equiv R_{\rho}(\v
x)=-i<\phib\p_x\phi>/g_2.\label{doper}\ee Through the $q_{\mu}\bar{\rho}\p_{\mu}\rho$ term in \Eq{totd2} this
gives rise to $ R_{\rho}(\v x)\bar{\rho}\p_{t}\rho.$ Aside from the spatial dependence of $R_{\rho}$ this looks
exactly like the Berry phase term in \Eq{phi}. If we view $\rho$ as  the phase factor of bosons and
\Eq{rho} as the Lagrangian for a system with integral number of bosons per unit cell, then the $\rho$
disordered phase corresponds to the bosonic Mott insulator. In this correspondence \Eq{doper} acts like a
spatial-dependent chemical potential. In regions where $|R_{\rho}(\v x)|$ exceeds the Mott gap an extra boson
density will be induced on the Mott insulating background. As the result a boson condensate puddles is nucleated.
Far away from the vortex $|R_{\rho}(\v x)|$ falls below the Mott gap consequently the Mott vacuum remains intact.
This means that stripes exist in the vicinity of a superconducting
vortex.
We note that in the above mechanism it is the screening
current of the superconductor that induces stripe order. This aspect is similar to Ref.\cite{sach1}. 

Finally we look at phase c) where there is stripe order without superconductivity. The lack of $\phi$ order requires
$R_{\phi}=0$. Thus the boson density satisfies the requirement that there is an integer number
of Cooper pairs per stripe unit cell.  In this phase if a dislocation is
nucleated somewhere, the phase of $\rho$ will wind by $2\pi$ around it. Through the same mechanism discussed in the previous
 paragraph a spatially varying $ <J_t(\v x,t)>=R_{\phi}(\v x)$ will be induced, which leads to superconducting puddles
near the dislocation.


In the presence of disorder we expect the existence  of a finite density of dislocations. 
The superconducting patches associated with these dislocations form a random granular array. It is possible that
with Josephson tunneling superconductivity can be sustained among these grains. 
In this
connection we noted that in a quite different context Zaanen and Nussinov made an interesting conjecture that the
high $T_c$ superconducting state corresponds to the condensation of topological excitations of an ordered stripe
state\cite{zz}.

In the above we have assumed that the charge density wave order spontaneously breaks the $90$ degree rotation
symmetry of the underlying lattice. It is interesting that removing this assumption has no impact on the duality
at all. After adding the orientation order parameter \Eq{tot} becomes \ba
\cL&&=\frac{K_{\rho\perp}}{2}|(\p_{\perp}-ig_1J_t)\rho|^2+ \frac{K_{\rho
t}}{2}|(\p_t-ig_2J_{\perp})\rho|^2\nn&&+\frac{K_{\rho\parallel}}{2}|\p_{\parallel}\rho|^2+\frac{1}{2K_{\phi
\parallel}}J_{\parallel}^2 + \frac{1}{2K_{\phi
\perp}}J_{\perp}^2 + \frac{1}{2K_{\phi t}}J_{t}^2 \nn&&+ J_{\parallel}\phib\p_{\parallel}\phi +
J_{\perp}\phib\p_{\perp}\phi + J_{t}\phib\p_{t}\phi +R_{\phi}\phib\p_t\phi\nn &&+ \frac{K_{\theta 1
}}{2}|\p_{x}\chi|^2+  \frac{K_{\theta 1 }}{2}|\p_{y}\chi|^2+ \frac{K_{\theta 2 }}{2}|\p_{t}\chi|^2
+V(\theta).\label{totot}\ea Here $\p_{\perp}=\cos\theta\p_x+\sin\theta\p_y$,
$\p_{\parallel}=\cos\theta\p_y-\sin\theta\p_x$ where $\theta$ is the angle between the stripe ordering direction
and the crystalline axis, $\chi=e^{i\theta}$ is the orientation order parameter, and $ V(\theta)=-K \cos{4\theta}$
acts to pin $\theta$ to $0,\pi/2,\pi,3\pi/2$. Starting with \Eq{totot} we can redo duality transformation between
$\rho$ and $\phi$ with $\chi$ unaffected.

Doped Mott insulator often exhibit inhomogeneous structure of competing orders. Unidirectional charge density wave
and superconductivity seems to compete in the LSCO family of the cuprates\cite{lake,app}. Similar competition
between a charge ordered state and ferromagnetism exists in the colossal magneto-resistive materials\cite{burgy}.
In this paper we present an effective field theory for competing superconducting and unidirectional charge density
wave orders\cite{note10}. We find an interesting duality\cite{comm} between these two types of order (Eqs.
\ref{tot},\ref{totd2},\ref{rep}). An implication of this duality is the presence of stripes near a superconducting
vortex, and the presence of superconductivity near a stripe dislocation. The evidence of the former is already
obtained in Ref.\cite{lake}. To check the latter one can apply
stress to stripe ordered state to induce dislocations. 



Acknowledgement: We are very grateful to Steven Kivelson and John Toner for helpful discussions. DHL is supported
in part by NSF grant DMR 99-71503.

\widetext


\begin{references}
\bibitem{nd} J.M. Tranquada {\it et al} Nature {\bf 375}, 561 (1995).
\bibitem{stripe} J. Zaanen, O. Gunnarsson Phys. Rev. B. {\bf40}, 7391 (1989); H.J. Schulz Phys. Rev. Lett.
{\bf64}, 1445 (1990); V.J. Emery, S.A. Kivelson Physica C {\bf209}, 597 (1993).
\bibitem{lake} B. Lake {\it et al} Science {\bf291}, 1759 (2001).
\bibitem{so5} A closely related idea where the stripe order is replaced by the commensurate antiferromagnetic order
can be found in the following references: S.-C. Zhang Science {\bf275}, 1089 (1997); D.P. Arovas {\it et al} Phys. Rev. Lett. {\bf 79}, 2871
(1997).
\bibitem{app}E. Demler, S. Sachdev, Y. Zhang Phys. Rev. Lett. {\bf87}, 067202 (2001); S.A. Kivelson, G. Aeppli,
V.J. Emery cond-mat/0105200; J.-P. Hu, S.C. Zhang, cond-mat/0108273.
\bibitem{ed} An interesting theory for harmonic fluctuating stripes and superconductivity
can be found in S.A. Kivelson, E. Fradkin, V.J. Emery Nature {\bf393}, 550 (1998).
\bibitem{lf} M. P.A. Fisher,  D-H Lee Phys. Rev. B {\bf 39}, 2756 (1989);
D-H Lee, M.P.A. Fisher Int. J. Mod. Phys. B {\bf5}, 2675 (1991).
\bibitem{nel} See D.R. Nelson, ``Defect-mediated Phase Transitions'', in {\it Phase transitions and critical phenomena vol 7}, Edited by C. Domb and J.L. Lebowitz.
\bibitem{monopole} The fact that commensurate pinning always pins the stripes is equivalent to the statement that
in the presence of magnetic monopole a compact U(1) gauge theory on a lattice is always in the
confining phase. See  A.M. Polyakov {\it Gauge Fields and
Strings} (Harwood Academic, Chur, 1987).
\bibitem{yamada} K. Yamada {\it et al}, Phys. Rev. B {\bf57}, 6165
(1998).
\bibitem{sach1} A. Polkovnikov {\it et al} cond-mat/0110329.
\bibitem{zz} J. Zaanen, Z.  Nussinov  NATO Science Series, II {\bf 15}, 129 (2001).
\bibitem{burgy} J. Burgy {\it et al}, Phys. Rev. Lett. {\bf 87}, 277202 (2001).
\bibitem{note10} This work should also apply to conventional S-wave superconductivity if it so happens that it
competes with an unidirectional charge density wave order.
\bibitem{comm}The duality discussed in this paper is
very different from the duality between the charge density wave and superconducting orders in one dimensional
systems. In the present work the stripe and the superconducting orders are characterized by two distinct order
parameters, while in the one dimensional duality the charge density wave and superconducting order parameters are
the dual representations of the same quantity. 

\end{references}
\end{document}